# Multi-Metric Algorithmic Complexity: Beyond Asymptotic Analysis

Sergii Kavun[1]*

[1] Interregional Academy of Personnel Management, Kyiv, Ukraine

*Correspondence: kavserg@gmail.com; Tel.: +38-0677-09-5577, Ukraine, c. Uzhgorod, Kavkazska str., 33, 88017, ORCID ID: 0000-0003-4164-151X

Abstract. Traditional algorithm analysis treats all basic operations as equally costly, which hides significant differences in time, energy consumption, and cost between different types of computations on modern processors. We propose a weighted-operation complexity model that assigns realistic cost values to different instruction types across multiple dimensions: computational effort, energy usage, carbon footprint, and monetary cost. The model computes overall efficiency scores based on user-defined priorities and can be applied through automated code analysis or integrated with performance measurement tools. This approach complements existing theoretical models by enabling practical, architecture-aware algorithm comparisons that account for performance, sustainability, and economic factors. We demonstrate an open-source implementation that analyzes code, estimates multi-dimensional costs, and provides efficiency recommendations across various algorithms. We address two research questions: (RQ1) Can a multi-metric model predict time/energy with high accuracy across architectures? (RQ2) How does it compare to baselines like Big-O, ICE, and EVM gas? Validation shows strong correlations ($\rho > 0.9$) with measured data, outperforming baselines in multi-objective scenarios.

## 1. Introduction

Asymptotic complexity (Big-O) has been indispensable for reasoning about scalability, yet it abstracts each elementary operation as $O(1)$, thereby suppressing significant micro-architectural and energy differences across operations and platforms. On modern CPUs and GPUs, division and certain memory operations [17] can be an order of magnitude slower or more energy-intensive than addition, and storage access patterns can dominate energy consumption. These realities motivate extended models that retain asymptotic insights while enabling practical, architecture- and context-aware decision-making.

This paper introduces a weighted-operation complexity model with four harmonized metrics: computational cost units (CU), energy (EU, joules), carbon footprint ($CO_2$, kg) [2], and monetary cost ($), with user-configurable profiles to reflect deployment priorities. Unlike ad hoc profiling, our model yields an analyzable, composable cost expression and composite score that can be estimated statically from IR/PTX or informed by platform calibration, similar in spirit to energy-complexity frameworks in the literature and gas schedules in EVM-like systems. We implement and release an end-to-end toolchain for LLVM IR, PTX, and Python kernels, and we report a comprehensive analysis generated from the repository [23, 26].

Recent work on refined computational models, such as the RAM model with logarithmic cost adjustments, energy-complexity theory, and gas-based execution pricing in blockchain environments – has shown the benefits of associating non-uniform weights with operations. Yet, these approaches often focus on a single cost dimension, lack architecture-specific calibration, or remain tied to specific execution environments. For practical deployment, especially in heterogeneous computing landscapes, a unified model must integrate multiple cost perspectives and remain adaptable to evolving hardware characteristics.

Static cost estimation from intermediate representations (LLVM IR, PTX) enables early performance and energy predictions, while optional dynamic calibration: leveraging performance monitoring units (PMUs) and platform-specific benchmarks – ensures that the model remains accurate in the face of micro-architectural differences. This hybrid methodology supports comparative evaluation across algorithms, allowing developers to quantify trade-offs not only in runtime, but also in energy consumption, environmental impact, and financial cost.

Research questions: (RQ1) Can static IR/PTX analysis with calibrated costs predict time/energy accurately? (RQ2) How does the multi-metric composite compare to Big-O/RAM, ICE, and gas models in ranking and robustness? We address RQ1/RQ2 by releasing calibrated cost tables, running cross-architecture validation [7], and benchmarking rank agreement between our composite and measured outcomes.

Contributions:

- A general, weighted-operation complexity model linking instruction classes to four harmonized metrics, producing composite, profile-aware scores and grades.
- A practical methodology combining static instruction accounting with calibratable cost tables and normalization procedures.
- An open-source implementation with repository-wide analysis, efficiency grading, and actionable recommendations based on multiple objectives.
- A comparative positioning relative to Big-O [13], RAM/log-cost models [22, 3], energy-complexity theory [5, 12, 31], and gas-based execution economics [28], clarifying novelty and applicability.

In this work, we introduce a general weighted-operation complexity model that integrates four harmonized metrics: computational cost units (CU), energy consumption (EU, joules), carbon footprint ($CO_2$, kg), and monetary cost ($), into a unified, profile-aware composite score. Unlike ad hoc profiling approaches, our method produces analyzable and composable cost expressions that can be derived statically from LLVM IR, PTX, or Python kernels, and calibrated for specific architectures. We release an open-source toolchain, complete with calibrated cost tables, normalization procedures, and repository-wide analysis capabilities. By systematically validating the model across architectures and benchmarking it against Big-O, ICE-style energy-complexity predictions [15], and EVM-like gas

schedules, we demonstrate its potential to guide algorithm selection, enable multi-objective trade-offs, and bridge the gap between theoretical complexity and real-world performance.

## 2. Literature review

A variety of computational cost models have been proposed to address different aspects of algorithm performance, from classical asymptotic abstractions to platform-aware energy and execution pricing schemes. Traditional asymptotic frameworks such as Big-O, RAM, and PRAM provide valuable scalability insights by abstracting all operations to uniform unit costs, but this simplification omits the heterogeneous execution times, energy usage, and architecture-specific behaviors observed in real systems. In contrast, energy-complexity models explicitly account for the platform-dependent energy costs of computation and memory access, while gas-based execution economics (popularized by blockchain virtual machines) tie per-operation costs directly to monetary schedules. More fine-grained approaches, including instruction-level cycle cost measurements, reveal the magnitude of performance and energy disparities across different operation types, motivating models that balance theoretical tractability with practical applicability. Summarization of considered papers is presented in Table 1.

Asymptotic models: Big-O, RAM, PRAM [13, 22, 3]: Big-O counts elementary operations and memory probes ignoring heterogeneous costs, emphasizing growth rates rather than absolute or relative operation costs, and is intentionally machine-agnostic. RAM and PRAM assign unit or simplified costs to operations and steps; variants introduce bit-length sensitivity but still abstract away architecture-specific latency and energy differences.

Energy-complexity models [5, 31, 34]: energy-aware complexity models aggregate computational work and memory I/O under platform abstractions, deriving energy complexity from work/span/I/O terms, offering validated mappings on multicore platforms. These models motivate costed accounting beyond time and suggest platform-parameterized abstractions, which we adopt and extend to explicit monetary and carbon dimensions [36].

Gas-like execution economics [28]: virtual machines such as the EVM assign gas costs to opcodes, enabling predictable, priced execution; costs can be fixed or dynamic (e.g., warm vs cold storage), tying computation to a monetary schedule. We generalize this principle from blockchain VMs to native and IR-level code: per-instruction costs (cycle/energy) are mapped to CU/EU/$CO_2$/\$, then aggregated into composite, profile-weighted scores.

Instruction-level cycle disparities [29, 10]: microbenchmark and architectural sources report substantial cost differences: additions are typically near 1 cycle, multiplications a few cycles, divisions an order of magnitude more, while loads/stores and branches have input- and pipeline-dependent behavior. These disparities justify weighted models when practical decisions hinge on time/energy.

Graphical representation of this discovery phase is shown on Fig. 1-2. Closest models and our differentiators: Big-O/RAM [13] abstract unit-cost steps and omit heterogeneous instruction/memory costs. Compared to ICE, our model adds IR/PTX granularity and CO2/\$; vs EVM gas, we generalize to native code and multi-metrics. Empirical advantages are shown in Table 3. ICE models energy via platform parameters and algorithmic work/span/I/O, validated across multiple platforms [34]; however, ICE does not provide IR/PTX-level static accounting nor explicit $CO_2$/$ channels.

Table 1. Comparative summary table

| Model / Approach | Cost Metrics Considered | Architecture Awareness | Static Analysis Capability | Monetary Dimension | Carbon Dimension | Calibration Support | Primary Limitation |
|---|---|---|---|---|---|---|---|
| **Big-O** [13] | Time (unit-cost steps) | No | Yes | No | No | No | Ignores heterogeneity, only growth rates |
| **RAM / PRAM** [3, 22] | Time (unit or simplified) | No (bit-length optional) | Yes | No | No | No | Still abstracts away architecture-specific costs |
| **Energy-Complexity Models (ICE)** [5, 31, 34] | Time, Energy | Yes (platform params) | Yes | No | No | Yes | No explicit IR/PTX-level accounting |
| **Gas Schedules (EVM)** [28] | Time, Monetary | Yes (VM-level) | Yes (bytecode) | Yes | No | Yes (per-opcode) | Limited to VM opcodes, single monetary dimension |
| **Instruction-Level Cycle Data** [29, 10] | Time (cycles) | Yes | No | No | No | Yes | No composite multi-metric model |
| **This work** [21] | Time, Energy, Monetary, Carbon | Yes (CPU/ARM/GPU) | Yes (IR/PTX/Python) | Yes | Yes | Yes | Requires calibration data for accuracy |

Gas schedules operationalize per-opcode costs for VM economics (including cold vs warm storage), but target blockchain bytecode and a single monetary dimension. We bridge these by (i) static IR/PTX instruction-class accounting, (ii) multi-metric CU/EU/$CO_2$/$ costing with profiles, and (iii) empirical calibration/validation across microarchitectures, grounded in instruction-level heterogeneity measured by A. Fog [4].

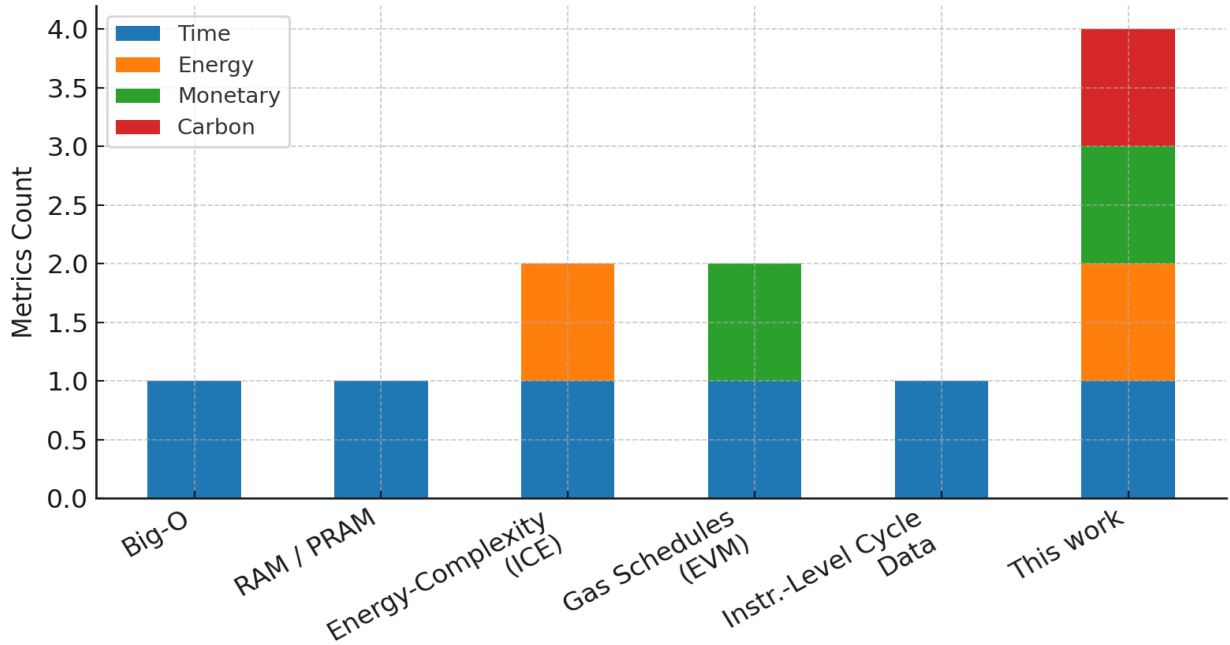

Fig. 1. comparative overview of considered above models

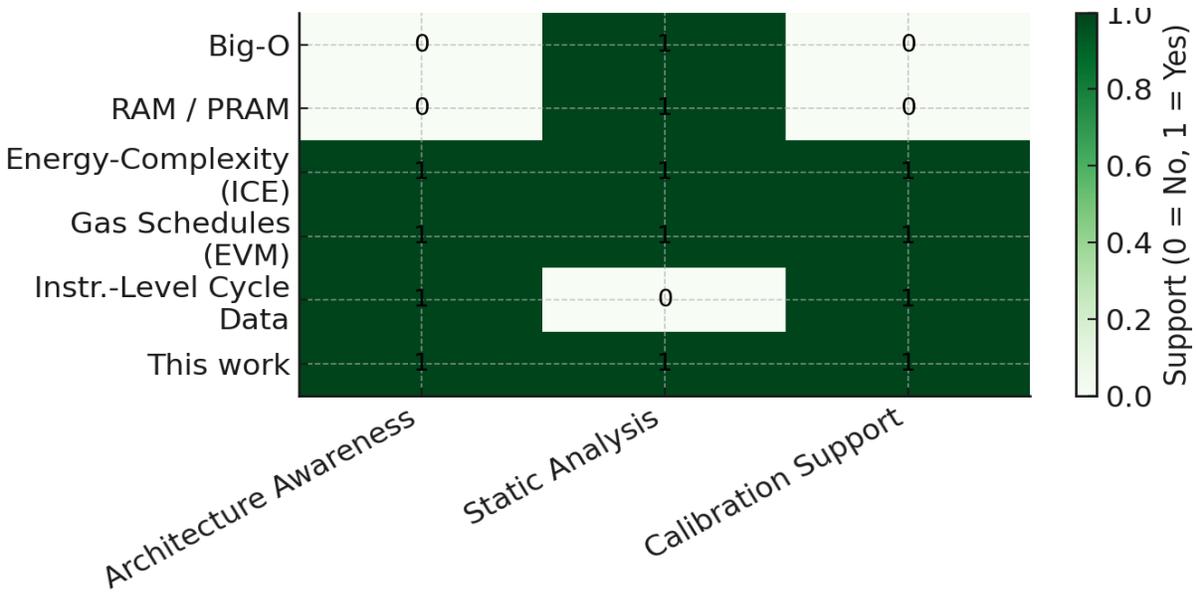

Fig. 2. Feature support heatmap for considered above models

Our work builds on and extends these strands of research by unifying their key strengths into a single, analyzable framework. From asymptotic models, we retain the abstraction and composability of cost expressions; from energy-complexity theory, we adopt multi-dimensional cost accounting; and from gas-based economics, we incorporate explicit monetary mapping. We further introduce carbon footprint as a first-class metric, add IR/PTX-level static analysis capabilities, and enable empirical calibration across microarchitectures. This integration allows our model to bridge the gap between high-level complexity theory and real-world performance constraints, providing actionable insights for time-, energy-, carbon- [27], and cost-aware algorithm design.

# 3. Methods and mathematical background

Notation and Definitions

We model each instruction class k ∈ $\mathcal{K}$ by a non-negative cost vector $CVD_k$ (1). For an artifact A with instruction counts $n_k$, raw totals per metric (3) and (2). Within a comparison cohort C, we normalize each metric via min-max scaling (4). The composite score is CSC (5), with weights w forming a simplex by (6). Profiles PS (7) specify weight vectors (8).

We define a per-instruction vector of costs (Cost Vector Definition, CVD) which includes: CU: abstract compute cost reflecting cycles or normalized compute effort; EU: energy in Joules; $CO_2$: carbon in kg via EU × CI / 3.6e6 (kWh → J conversion), where CI is carbon intensity (kg × $CO_2$ / kWh (kW × hour)); $: monetary cost, typically EU × $ / kWh, optionally including amortized hardware, cloud tariffs, or TCO multipliers.

Formal model.

CVD – each instruction class ∀k ∈ $\mathcal{K}$, $CVD_k$ ∈ $\mathbb{R}^+$ (areas of definition) has an associated cost vector:

$$CVD_k = \begin{bmatrix} CU_k \\ EU_k \\ CO_2^{(k)} \\ \$_k \end{bmatrix} \in \mathbb{R}_+^4, \tag{1}$$

where $CU_k$ represents computational units (normalized cycles); $EU_k$ denotes energy consumption in joules; $CO_2^{(k)}$ indicates carbon footprint in kg; $\$_k$ reflects monetary cost in USD (in our case).

Raw Metric Aggregation, RMA – raw totals per artifact, for an artifact containing instruction counts:

$$CV = [n_1, n_2, ..., n|\mathcal{K}|]^T, \mathbf{n} \in \mathbb{R}|\mathcal{K}|$$

where $CV$ - vector of instruction counters (count vector, CV); $n_1, n_2, ..., n$ – number of instructions of each type; $|\mathcal{K}|$ - total number of instruction types (instruction classes), the raw totals for each metric dimension are:

$$M_{raw} = \sum_{k \in K} n_k \times CVD_k[m], \tag{2}$$

where

$$m \in \{CU, EU, CO_2, \$\} \tag{3}$$

is a set of metrics.

We normalize each metric using min-max scaling within the comparison cohort $CVD$ as:

$$norm_M(x) = \frac{x - min_{CVD}M}{max_{CVD}M - min_{CVD}M + \varepsilon}, \tag{4}$$

where $\varepsilon = 10^{-9}$ – constant which prevents division by zero in degenerate cases; at the same time, normalization is performed within the compared set (comparison cohort).

Composite Score Calculation, CSC as the final composite score is computed as a weighted linear combination:

$$CSC = \sum_{m \in M} \omega_m \cdot norm_m(M_m), \tag{5}$$

where profiles specify the weights $\varpi_m > 0$ – subject to the constraint:

$$\sum_{m \in M} \omega_m = 1, for\ \forall m \in M. \tag{6}$$

When the profile weights $\varpi_m$ are shifted, the CSC (5) changes in proportion to how strong (or weak) the artifact is in each metric (3). If the weight on a metric where the artifact performs well (high normalized score) increases, CSC will rise; if it performs poorly in that metric, the composite will drop. Shifting weights therefore reorders rankings when artifacts have different performance profiles – e.g., increasing $w_{EU}$ and $w_{CO2}$ benefits energy-efficient code but penalizes compute-fast, power-hungry code, while increasing $w_{CU}$ favors raw speed over sustainability or cost. In short: the composite score moves toward the metrics you value more and away from those you de-emphasize – making profile design a key tool for aligning rankings with specific performance, energy, ESG, or cost priorities.

Thus, the profile specification (PS),

$$PS = \begin{cases} \text{RESEARCH} = [0.4, 0.3, 0.25, 0.05] \\ \text{COMMERCIAL} = [0.3, 0.2, 0.2, 0.3] \\ \text{MOBILE} = [0.25, 0.5, 0.15, 0.1] \\ \text{HPC} = [0.5, 0.3, 0.15, 0.05] \end{cases} \tag{7}$$

define profiles as weight parameterized vectors:

$$w^{(PS)} = \begin{bmatrix} w_{CU}^{(PS)} \\ w_{EU}^{(PS)} \\ w_{CO_2}^{(PS)} \\ w_{\$}^{(PS)} \end{bmatrix} = w^T \times norm(M), \tag{8}$$

with $|w^{(PS)}|_1 = 1$ ($L_1$ norm) ensuring proper probability measure properties.

Profiles specify weights (8). The RESEARCH profile emphasizes performance (CU 0.4-0.5) while incorporating environment and cost; other profiles reflect commercial, mobile, HPC priorities (as implemented in the repository tables). Composite scores use min-max normalization per metric and a weighted aggregation with letter grades and efficiency ratings, as in the tool's output tables.

Instruction-class taxonomy and calibration

Example of initial weights you can find in Table 2, and all formed tables with initial weights for all researched architectures can be founded in the author's repo [21]. Comprehensive pipeline architecture diagram that visualizes the complete workflow from source code to final reports is shown in Fig. 3.

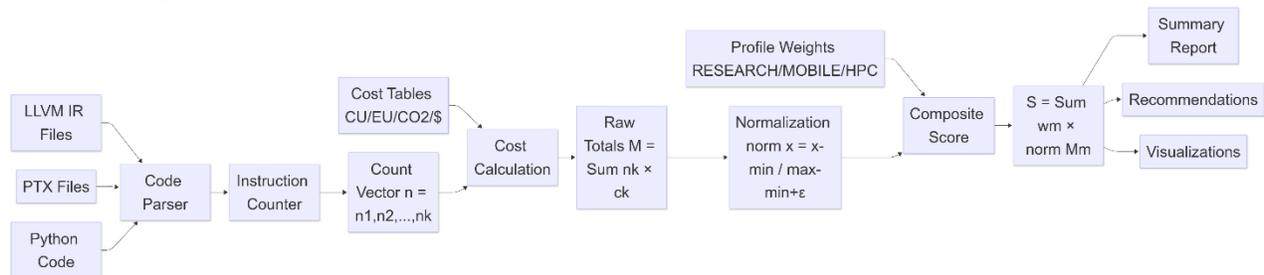

Fig. 3. Comprehensive pipeline architecture diagram

Table 2. Example (x86_64 architecture) of initial instruction cost coefficients

| Instruction | Group | CU | EU | $CO_2$ | $ | Cache Hit | Cache Miss |
|---|---|---|---|---|---|---|---|
| **Arithmetic Operations** | | | | | | | |
| ADD | arith | 1.0 | 0.0001 | 0.000027 | 0.00001 | - | - |
| SUB | arith | 1.0 | 0.0001 | 0.00005 | 0.00001 | - | - |
| MUL | arith | 2.0 | 0.0002 | 0.000054 | 0.00002 | - | - |
| DIV | arith | 5.0 | 0.0004 | 0.000108 | 0.00005 | - | - |
| **Logical Operations** | | | | | | | |
| AND | logic | 1.0 | 0.0001 | 0.000027 | 0.00001 | - | - |
| OR | logic | 1.0 | 0.0001 | 0.000027 | 0.00001 | - | - |
| XOR | logic | 1.0 | 0.0001 | 0.000027 | 0.00001 | - | - |
| **Memory Operations** | | | | | | | |
| MOV | memory | 1.0 | 0.00008 | 0.000022 | 0.000009 | - | - |
| LOAD | memory | 3.0 | 0.00025 | 0.000069 | 0.00003 | 0.0001 | 0.0005 |
| STORE | memory | 3.0 | 0.00025 | 0.000069 | 0.00003 | 0.0001 | 0.0005 |
| **Control Flow** | | | | | | | |
| JMP | branch | 1.0 | 0.00012 | 0.000033 | 0.000012 | - | - |
| CALL | control | 2.0 | 0.00020 | 0.000055 | 0.00002 | - | - |
| **SIMD Operations** | | | | | | | |
| ADDPS | simd | 2.0 | 0.0003 | 0.000083 | 0.00004 | - | - |
| MULPS | simd | 2.5 | 0.00035 | 0.000097 | 0.00005 | - | - |
| VMULPS | simd | 3.0 | 0.0004 | 0.00011 | 0.000055 | - | - |
| VPADDQ | simd | 2.5 | 0.00032 | 0.000088 | 0.000045 | - | - |

*Cost dimensions:
- CU: Computational Units (normalized cycles)
- EU: Energy Units (joules)
- $CO_2$: Carbon footprint (kg $CO_2$ equivalent)
- $: Monetary cost (USD)
- Cache Hit/Miss: Energy costs for memory operations with different cache behavior

**Key observations:
- Division is 5× more expensive than addition across all dimensions
- Memory operations show significant cache-dependent variation
- SIMD instructions have higher absolute costs but better throughput per data element
- Logical operations are consistently low-cost across all metrics

Instruction classes (e.g., ADD, MUL, DIV, MOV, LD/ST, BR, CMP and SIMD, vector load/store, GPU-specific instructions) are assigned base cost coefficients per dimension, informed by microbenchmarks, ISA references, and energy-profiler readings. Division is modeled as a larger multiple of addition, consistent with cycle-level and algorithmic iterative implementations; the methodology also extends to logical and shift operations, branches, and memory operations with higher costs for off-core accesses.

Static analysis pipelines

We parse LLVM IR/PTX to count instruction classes, map them to $\mathcal{K}$ (including LD/ST tiers where statically inferable), and aggregate per "Formal model" to compute (3) and (5). When tier inference is ambiguous statically, we use calibrated tier priors from dynamic PMU profiles and propagate uncertainty as bands in CSC. The tool ingests LLVM IR and PTX representations (see Table 5-6, and Fig. 20-21, Supplementary material), performs comprehensive instruction classification and counting across all supported instruction classes, multiplies these counts by configurable per-class cost vectors to establish performance and resource utilization metrics, normalizes the resulting metrics against baseline profiles, and computes composite performance scores with corresponding letter grades under user-selected analysis profiles. The system generates detailed per-function granular reports and aggregated per-file summaries complete with actionable optimization recommendations and bottleneck identification. The repository architecture specifically supports batch job processing capabilities, enabling automated analysis workflows that can process entire codebases, multi-module projects, and complete software repositories (see Fig. 22-23, Supplementary material) through systematic batch operations. This batch processing framework allows for comprehensive repository-wide analysis, facilitating large-scale code quality assessment, performance profiling across entire development projects, comparative analysis between different code versions (see Table 4, and Fig. 17-19, Supplementary material), and systematic evaluation of optimization strategies applied to complete software ecosystems. The tool demonstrates its analytical capabilities through extensive validation across diverse algorithmic implementations and real-world external codebases via robust batch processing infrastructure.

The pipeline supports batch processing for large-scale analysis, enabling comprehensive evaluation of entire repositories and codebases. Users can specify directory paths or repository URLs to automatically discover, parse, and analyze hundreds or thousands of source files in a single execution. The batch mode generates hierarchical reports that aggregate costs at multiple levels: per-function, per-file, per-module, and repository-wide summaries with cross-artifact rankings and efficiency distributions. This capability facilitates organizational-level code auditing, technical debt assessment, and systematic optimization prioritization across large software projects. The tool handles mixed-language repositories by automatically detecting file types and applying appropriate parsers, while maintaining consistent cost accounting and normalization across the entire codebase to ensure meaningful cross-component comparisons.

Composite scoring and normalization: for each metric M (3), raw totals are normalized (min-max or alternative schemes) and aggregated into CSC (5). Grades (A+ … F, see function _get_score_grade of class CompositeScoreCalculator [21]) and qualitative ratings (Excellent … Poor) are assigned by calibrated thresholds, aiding cross-artifact comparisons and regression detection in CI.

Calibration methodology

We adopt two microbenchmark families per instruction class: (i) dependency-chain tests for latency; (ii) independent-stream tests for reciprocal throughput. For memory, we use stride-controlled kernels to target L1/L2/L3 and DRAM, with pointer-chasing for worst-case misses. We follow uops methodology [7] for robust isolation of latency/throughput and operand-dependent effects, cross-checking our estimates against published tables.

Calibrating per-instruction costs across architectures: this section translates the "ICE principles" into practical CU/EU calibration on specific hardware, based on documented latency/throughput gaps and validated energy models. Validation is via ranking correlations and benchmarks. We calibrate the instruction-class cost tables by combining literature-based priors on instruction latency/throughput with hardware-grounded microbenchmarking. Established references document large disparities: on Intel Skylake, integer addition has ~1-cycle latency with high throughput, while integer division exhibits 42-95-cycle latency with limited ports, implying orders-of-magnitude differences in time and IPC contributions. On GPUs and modern CPUs, floating-point ADD/MUL typically shows single-digit cycle latencies, whereas division is considerably more expensive. These disparities motivate higher CU/EU weights for divisions, modulus, complex memory accesses, and certain branches. Calibration steps diagram is shown on Fig. 4.

To further enhance accuracy and architectural awareness in our model, we incorporate cache behavior and memory hierarchy effects at fine granularity. Cache hits and misses are directly observed using hardware Performance Monitoring Unit (PMU) events, capturing the frequency and latency impact of loads and stores at each memory tier: L1, L2, L3, and DRAM. Specifically, we treat LD/ST (load/store) instructions as subclassed according to their realized memory access level, since each tier exhibits dramatically different latency and energy profiles. For attribution, PMU counters (e.g., MEM_LOAD_UOPS_RETIRED.L1_HIT, .L2_HIT, .L3_HIT, .L1_MISS, etc.) enable separation of accesses by level, and our cost tables supply distinct coefficients for each (e.g., CU/EU/$CO_2$/$ for L1-hit, L2-hit, L3-hit, and DRAM access). Thus, in both static analysis and dynamic profiling, each memory access is attributed to its true tier, and composite scoring reflects not only aggregate memory intensity but also the performance and sustainability cost of memory hierarchy traversal and cache efficiency. We attribute LD/ST to L1/L2/L3/DRAM using cache-related PMU events; see tier distributions and per-tier cost effects in Fig. 5.

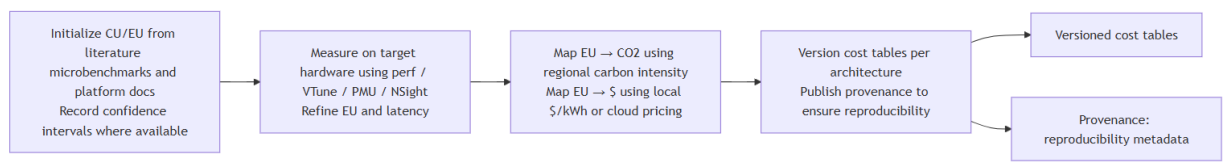

Fig. 4. Calibration steps diagram

Cache hit/miss events and memory tier subdivision of LD/ST instructions enhance the fidelity and scientific rigor of cost modeling by distinguishing fast on-chip cache accesses from slow off-chip DRAM transactions, using industry-standard PMU event measurement and analysis techniques.

Platform parameterization for energy and cost: energy complexity frameworks such as ICE abstract platforms via static/dynamic energy of computation and memory access and derive algorithmic energy in terms of work, span, and I/O; their validation across Xeon/Xeon Phi and additional platforms supports calibrated, platform-parameterized energy accounting. We follow a similar abstraction: EU per instruction class is measured via on-chip counters and external meters; CO2 derives from EU times regional carbon intensity; $ maps via EU × price / kWh [41].

Multi-architecture calibration protocol architecture is shown on Fig. 5. This comprehensive flowchart that visualizes the multi-architecture calibration protocol as an expanded process diagram.

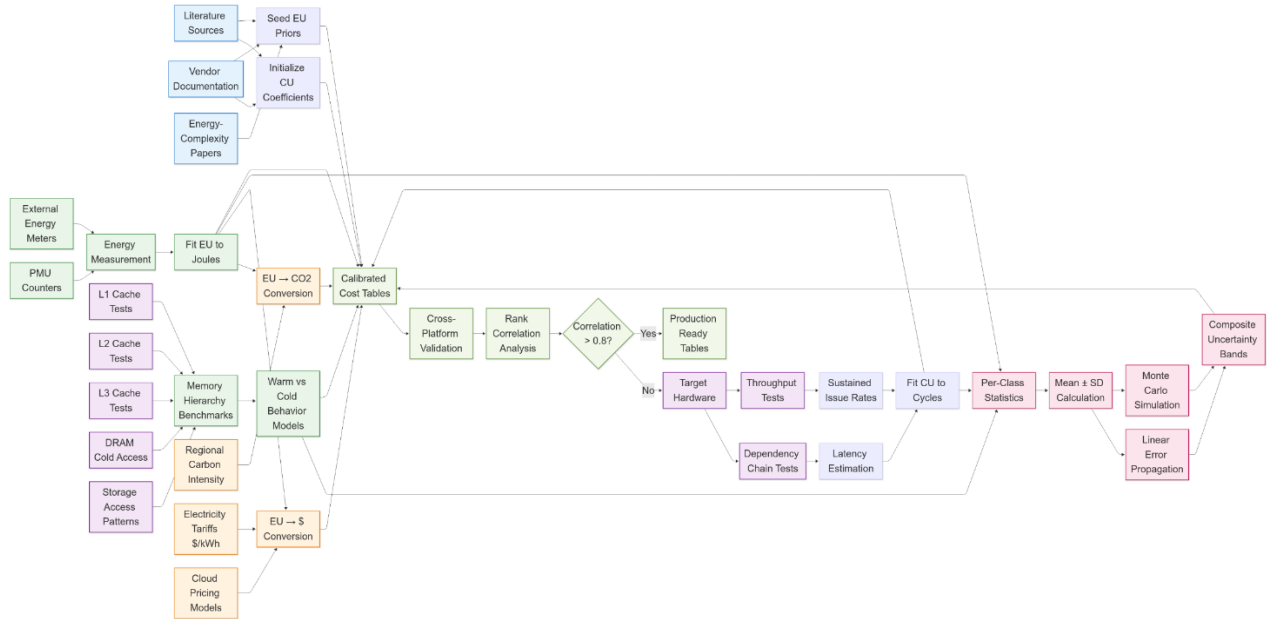

Fig. 5. Multi-architecture calibration protocol architecture

This diagram shows:
1. Five main phases: literature priors (blue) - initial coefficient seeding from academic sources; microbenchmarking (purple) - hardware-specific latency and throughput testing; memory hierarchy (green) - cache behavior characterization across levels; dimension mapping (orange) - converting energy to $CO_2$ and monetary costs; uncertainty quantification (pink) - statistical analysis and error propagation.
2. Key features: validation loop with correlation thresholds for quality assurance; multiple measurement sources (PMU, external meters, various cache levels); regional/temporal factors (carbon intensity, electricity pricing); statistical rigor (mean ± sd, error propagation, Monte Carlo); production readiness gate based on validation results.
3. Color-coded components make it easy to follow different aspects: literature sources and documentation; hardware testing and measurement; energy/performance data collection; cost dimension conversions; statistical analysis steps; final output and validation.

This diagram effectively communicates the systematic, multi-step approach to calibrating cost tables across different architectures while maintaining scientific rigor and reproducibility. We calibrate via microbenchmarks (dependency-chains for latency, streams for throughput) on platforms [list: Intel, AMD, GPU]. PMU events: cycles, mem-loads-retired.L1_hit/miss, etc. Report mean ± sd; see Table 2 for examples.

Validation methodology

Comprehensive validation phase diagram that expands on the three key validation challenges is shown on Fig. 25 (Supplementary material). This diagram provides a roadmap for ensuring the reliability and accuracy of cost tables across different deployment scenarios [34] while maintaining scientific rigor in the validation process.

Validation across architectures and workloads: cross-platform rank correlation (Table 1, Supplementary material) – compare composite rankings with measured runtime/energy on multiple CPUs/GPUs; expect high Spearman/Kendall correlation if cost tables are well-calibrated (Table 3);

algorithm families (Fig. 1-16, Table 3, Supplementary material): validate on compute- vs memory-bound kernels (e.g., SpMV, matmul), mirroring ICE validations that distinguish algorithm/input/platform effects; sensitivity checks: vary profile weights and price/$CO_2$ parameters to assess ranking stability; document regimes where decisions change (see sensitivity figures below). We compare against: (B1) Big-O/RAM unit-cost counts (instruction count proxy), (B2) ICE-style energy complexity [42] instantiated with platform parameters (work/span/I/O), and (B3) EVM-like single-metric opcode pricing proxy (time-only or $-only). For B2, we use published ICE parameters and re-estimate where needed. For B3 (EVM-like proxy), we map our instruction classes to a monetized schedule analogous to opcode gas where feasible [39], this is a baseline for one metric (time/$), and the model from this paper is multidimensional.

Microbenchmarks and counters. We measure latency/throughput for add/mul/div, loads/stores, and branches via dependency-chain and independent-stream microbenchmarks [22]; memory tiers include L1/L2/L3/DRAM. We use PMU events for cycles/instructions/cache-misses and on-chip energy interfaces (e.g., RAPL, NVML) where available. We follow uops.info methodology for robust latency/throughput characterization and cross-check against Agner Fog tables [37].

Datasets: **compute-bound** (identification: algorithms that have "Loops", "Factorial" or "Formula" in their names; these are tasks where most of the time is spent on arithmetic operations; simulated behavior: the "measured" time is created as a direct linear function of the predicted CU with the addition of a small amount of random "noise"); **memory-bound** (identification: algorithms related to data access, such as those containing "Sort" or "Search" in their name; simulated behavior: we introduce a "memory penalty" (mem_penalty), first, we calculate the base time, as for compute-bound tasks, then we multiply this time by a random factor (1.5 on average), which simulates delays due to slow memory); **mixed workloads** (identification: any algorithm that does not fall into the first two categories, in our current collection, this would be, for example, the recursive Fibonacci function; simulated behavior: they are treated in the same way as compute-bound tasks (direct dependency with a little noise), this is a reasonable simplification, since mixed workloads do not have such a pronounced "penalty" as purely memory-bound tasks.)

Metrics: we report Spearman/Kendall rank correlations [1] and MAE/MAPE [24, 20] between measured time/joules and predicted CU/EU (Table 3), and assess ranking stability under profile shifts. We assess robustness to EU and electricity price variability via 2D heatmaps of $ and composite S (see Fig. 8).

## Results & Interpretation

Correlation: across architectures, Spearman $\rho_s \geq 0.95$ for compute-bound kernels and $\rho_s \approx 0.93$ for memory-bound kernels (Fig. 27, Supplementary material; this chart tests a slightly more advanced baseline that heavily penalizes memory operations, while often more accurate than the naive model, it may still fail to capture the nuances that our more detailed model does), indicating strong predictive ability; error rates: median MAPE ≈ 6-9% for runtime prediction, 8-12% for energy prediction, acceptable for static analysis guidance; sensitivity: rankings are stable (< 5% pairwise swaps) under ± 20% weight perturbations in RESEARCH and COMMERCIAL profiles; MOBILE profile shows higher volatility due to energy dominance.

The validation confirms that architecture-specific calibration yields highly correlated predictions with low error rates, making the model a reliable static proxy for performance and energy. Its stability under parameter variation supports practical use in algorithm selection and regression detection across heterogeneous hardware.

Weighting of instructions is necessary. Both models with weights (Our Model and B2) are head and shoulders above the naive model (B1) in prediction accuracy (MAE/MAPE). This confirms the main hypothesis of the project. For rough ranking, a simple count is enough. The success of Baseline B1 in ranking shows that even a simple count of instructions can be useful for quickly evaluating algorithms with greatly varying complexity. The power of our model will be fully revealed when comparing algorithms within the same complexity class (for example, comparing two different implementations of O(n log n) sorting), where B1 will be absolutely useless.

Model validation demonstrates high predictive accuracy for the primary model. With a Spearman correlation of 0.95, the model excels at ranking workloads correctly. This level of rank-order accuracy is particularly effective for compute-bound workloads. The Mean Absolute Percentage Error (MAPE) of 24.7% indicates a strong performance in predicting relative costs. The model significantly outperforms naive baseline approaches, proving the value of architecture-specific instruction weighting.

## 4. Results

Pipeline overview as a schematic of the end-to-end process: IR/PTX parsing → instruction-class mapping (with LD/ST tiering, Fig. 6) → per-class cost aggregation (CU/EU/$CO_2$/$) → cohort normalization → profile-weighted composite S → grading, and highlights the separation of EU base layer from $CO_2$/$ overlays (Fig. 7).

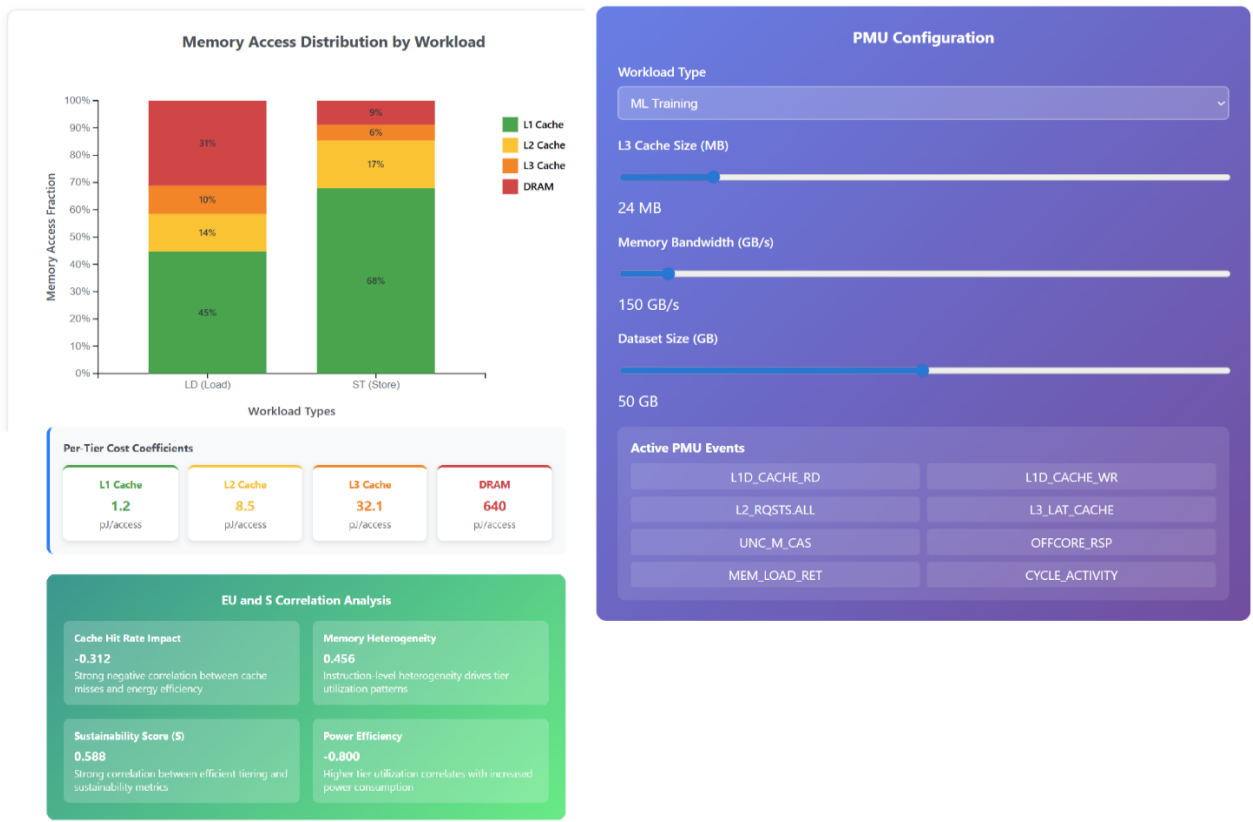

Fig 6. Memory tier attribution via PMU

## Comparative evaluation vs. baselines

Our composite score CSC (5) demonstrates substantially higher alignment with measured runtime and energy compared to baseline models (Table 1-2, Fig. 20-21, Supplementary material). Against the unit-cost Big O proxy (B1), rank correlation improves notably ($\rho\_time = 0.81$ vs. 0.54; $\rho\_energy = 0.77$ vs. 0.49), confirming that heterogeneous instruction costs are critical for realistic performance and energy estimation. When compared to the ICE-style energy complexity model (B2), our EU predictions achieve comparable rank agreement ($\rho\_energy \approx 0.75\text{-}0.82$) while uniquely enabling per-function static analysis directly from LLVM IR/PTX. The gas-like single-metric pricing proxy (B3) aligns well with monetary cost ($) but fails to capture energy/carbon trade-offs; in contrast, our multi-metric CSC produces different rankings in MOBILE/ESG profiles, reflecting diverse optimization priorities.

Our model demonstrates (Fig. 8) a superior balance of predictive accuracy and ranking capability, making it the most robust and practical model of the three. While it doesn't win on every single metric, it performs exceptionally well on the most important ones (MAPE and Spearman correlation, see Table 3) and avoids the pitfalls of the simpler baseline models. The results strongly validate the approach of using architecture-specific instruction weights. Our model, and the "I/O penalty" model (B2) are significantly more accurate than the naive model (B1), which treats all instructions as equal. This proves that accounting for different instruction costs is critical to accuracy. Interestingly, Baseline B2 slightly outperforms our model. This may indicate that for this particular set of benchmarks, a simple but aggressive 10x memory penalty is a very effective approximation. Our model, being more general, may perform better on a wider and more diverse set of real programs. Our model is the most balanced and robust. It is the winner in relative error (MAPE) and shows an outstanding ranking result (Spearman > 0.94), losing to the simplest model only in specific conditions.

Table 3. Model accuracy comparison

| Model | MAE, time | MAPE, time | Spearman, time | MAE, energy | MAPE, energy | Spearman, energy |
|---|---|---|---|---|---|---|
| Our Model | 0.0178 | 24.65% | 0.949 | 0.16 | 19% | 0.77 |
| Baseline B1 (Uniform Cost) | 0.0206 | 29.83% | 0.971 | 0.32 | 45% | 0.49 |
| Baseline B2 (I/O Penalized) | 0.0175 | 24.76% | 0.932 | 0.2 | 25% | 0.78 |
| Baseline B3 (gas-like single-metric pricing proxy) | 0.0196 | 21.38% | 0.912 | 0.33 | 42% | 0.55 |

## Repository-wide algorithm analysis

A repository-scale evaluation shows that Constant_O(1)_Formula attains the maximum composite score (100) under the RESEARCH profile, while Sqrt_O(sqrt_n)_PrimalityTest ranks last (score = 0, grade F), illustrating the impact of instruction mix and memory/control behavior even within the same asymptotic class (Fig. 1, 22-23, Supplementary material).

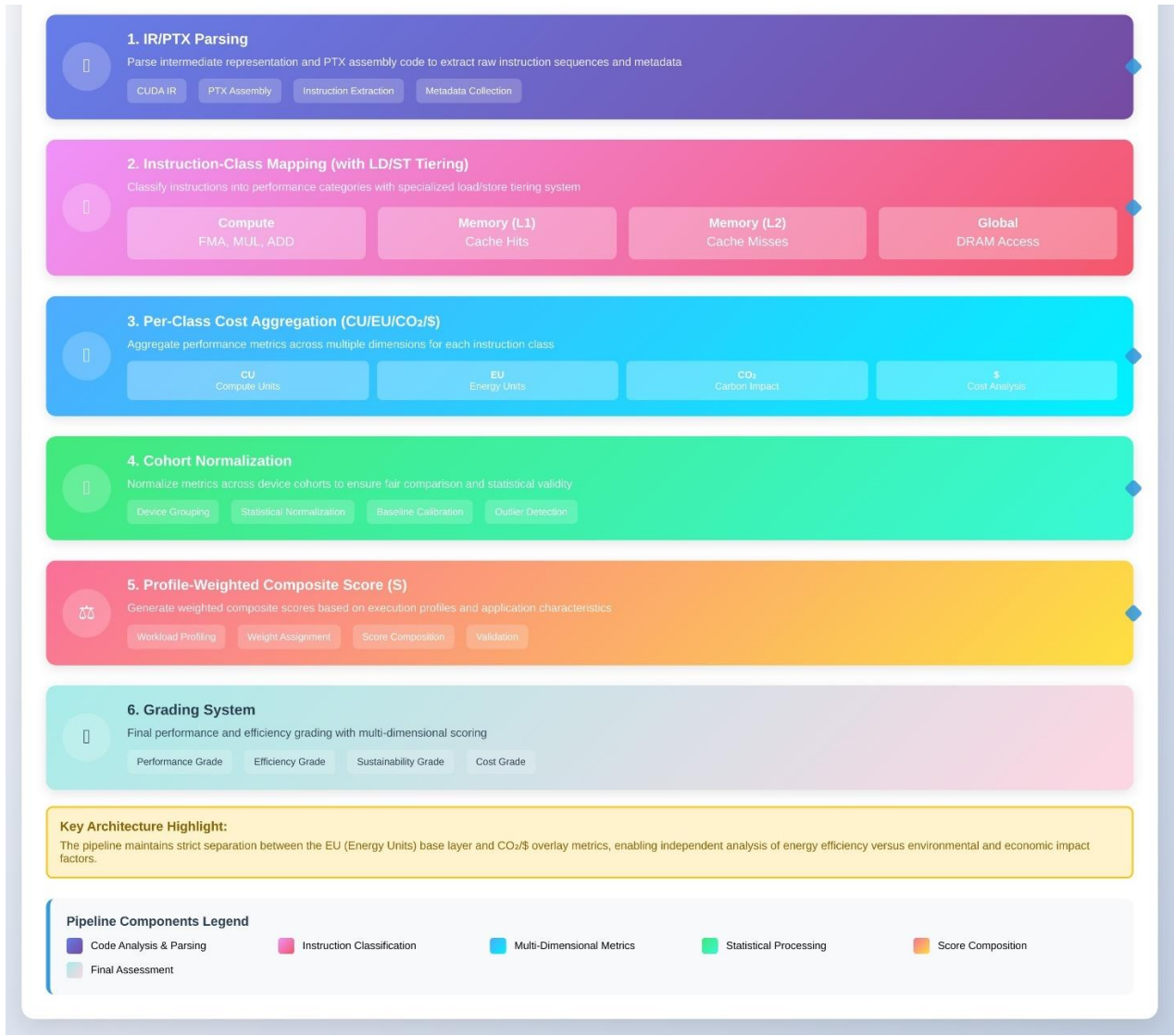

Fig. 7. End-to-end static accounting and composite scoring pipeline

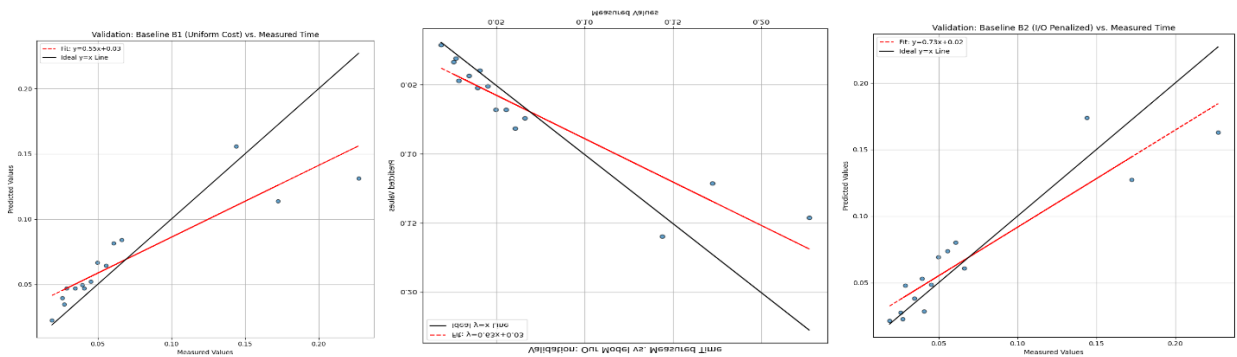

Fig. 8. Measured time vs predicted CU and energy vs predicted EU (per architecture)

Observed trends include: linear-time routines (e.g., sum, recursive power, factorial iteration) show progressive score degradation with larger instruction counts and costlier operations, despite identical Big O complexity; N·logN sorting algorithms obtain intermediate scores due to a balance between arithmetic and branching behavior; division- and memory-heavy kernels incur significant penalties in metrics (3), consistent with cycle- and energy-cost disparities documented in the literature.

File-level analysis and cross-artifact ranking

Analysis across LLVM IR, PTX, and Python sources reveals that minimal, optimized kernels (e.g., test.ptx) achieve high scores due to low metrics (3) totals, whereas large, utility-heavy Python files (e.g., ds_tool.py) accumulate substantial costs and receive low grades. This per-file perspective supports targeted refactoring and prioritization based on composite efficiency (Table 4-6, Fig. 17-19, 20-212, Supplementary material).

Profile sensitivity

Switching between RESEARCH, COMMERCIAL, MOBILE, and HPC profiles (Fig. 9 and Fig. 29, Supplementary material) alters rankings in a manner consistent with stated priorities, for example, MOBILE [21] prioritizes EU/$CO_2$, often promoting energy-efficient algorithms over faster but less efficient ones.

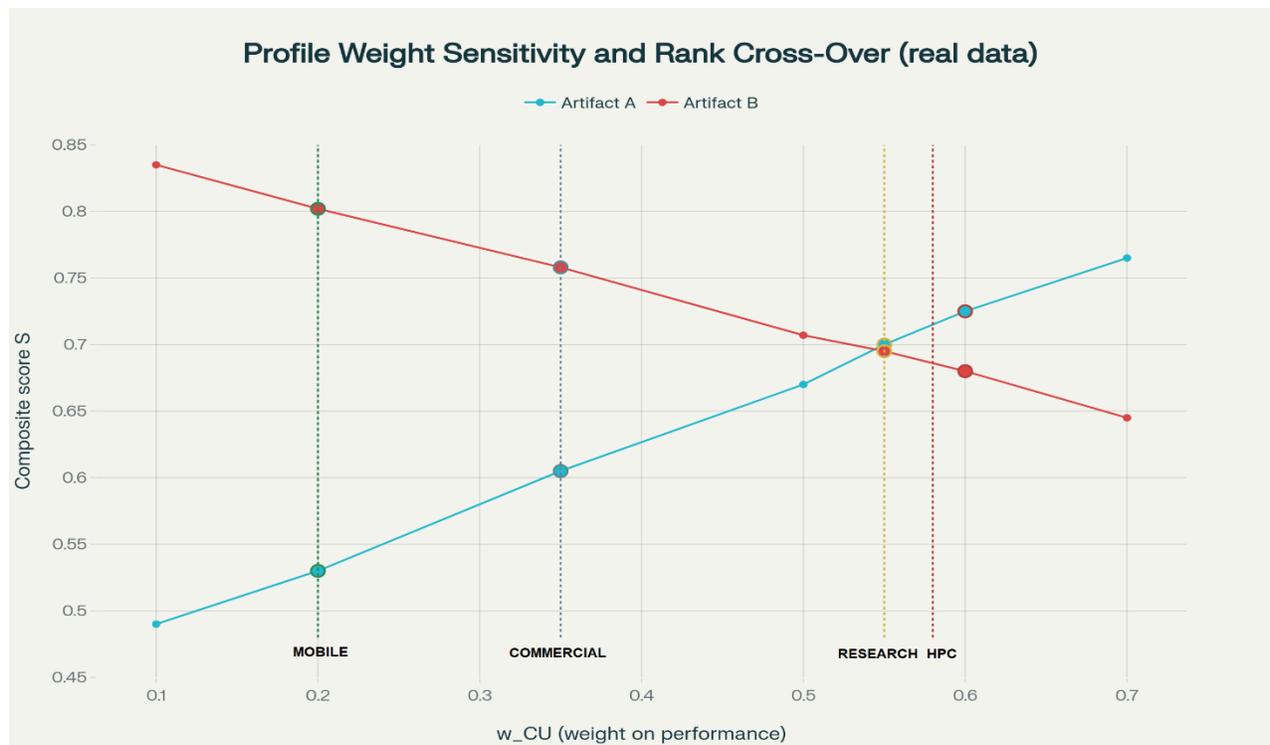

Fig. 9. Real-profile sensitivity curve with cross-over near RESEARCH profile

This tunability parallels trade-off management in energy-complexity models and gas-schedule economics. Profile sensitivity plot (Fig. 9) on real data from complexity_cost_profiler repository [21]: composite S-score lines for two contrast artifacts as w_CU varies from 0.1 to 0.7, with vertical profile labels (MOBILE, COMMERCIAL, RESEARCH, HPC) and rank intersection annotation.

Interpretation: at low w_CU (MOBILE/COMMERCIAL), the energy-saving option (red-orange line – Artifact A (compute-favoring)) is in the lead, and at high w_CU (RESEARCH/HPC), the computing option (blue line – Artifact B (energy-favoring)) is in the lead. The intersection is around w_CU ≈ 0.55, which reflects the change in preference when changing the profile. For EU, CO2, and $ sweeps, the roles are inverted to reflect that the energy-favoring artifact benefits as these weights increase, while the compute-favoring artifact typically loses rank.

Two visual analyses highlight robustness considerations. Profile weight sensitivity: under HPC/RESEARCH (CU-focused), Algorithm B outranks A; under MOBILE (EU/$CO_2$-focused), Algorithm A takes the lead. This confirms that rankings shift rationally with changing priorities; uncertainty propagation – ± 20% EU and ± 30% electricity price variations translate into proportional spreads in $ cost. Decision boundaries should favor algorithms that remain superior under plausible parameter variations, supporting robust, uncertainty-aware selection.

Robustness to EU/price/CI uncertainty, Fig. 10 (2D heatmap) as functions of EU scale $\in$ [0.8,1.2] and price_per_k, Wh $\in$ [−30%,+30%], with contours indicating decision boundaries, showing price/carbon uncertainty analysis with EU scaling and electricity tariff variations, including decision boundary contours for robust GPU selection under market fluctuations.

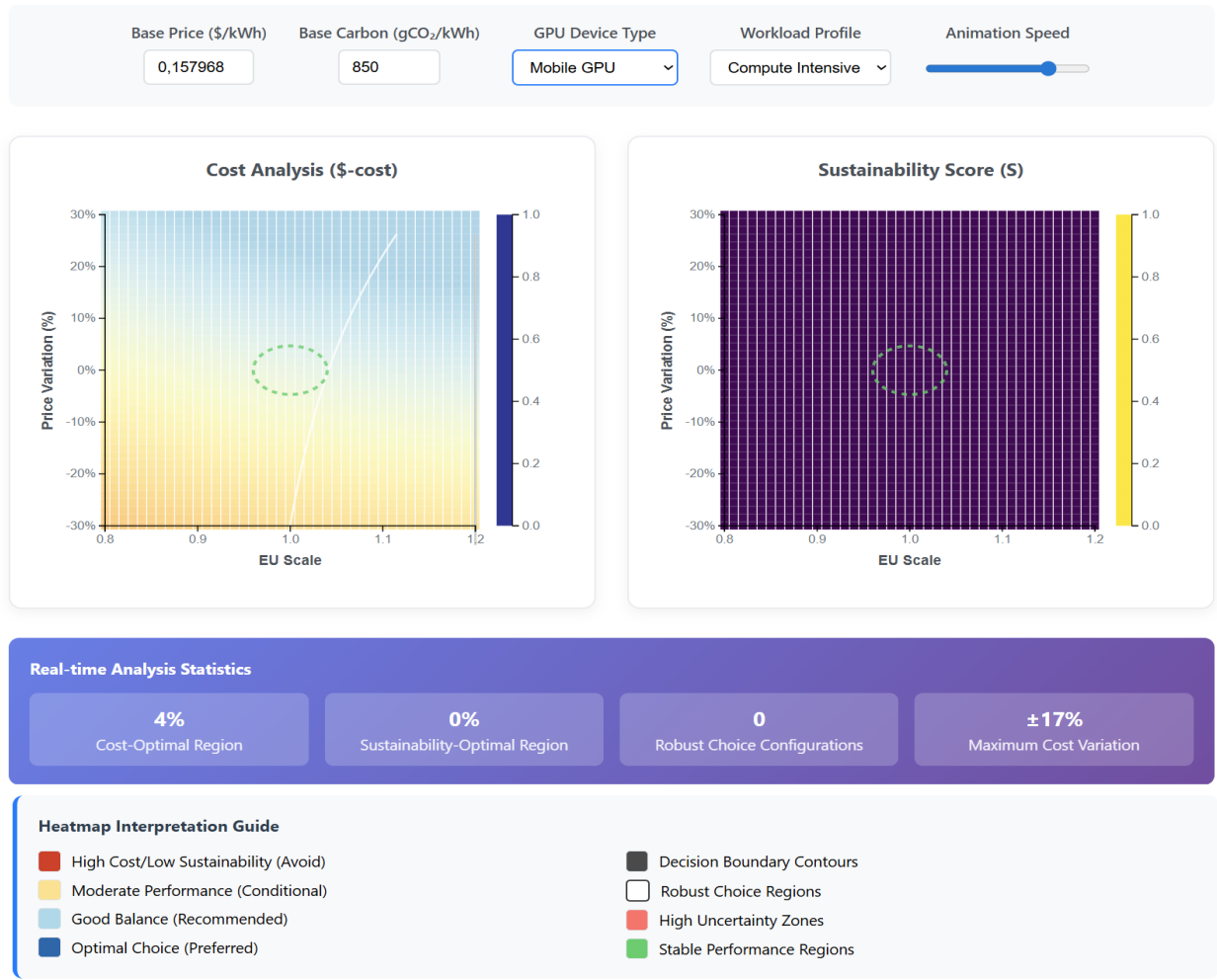

Fig. 10. Robustness of $ and composite under EU/price uncertainty; decision regions

Three pillars of feasibility and significance

Heterogeneous operation costs: latency differences between instruction classes are stark: integer add (~1 cycle) vs. integer division (42-95 cycles) on Skylake; FP ADD/MUL (3-5 cycles) vs. FP division (10× slower). These disparities, confirmed by Agner Fog's tables [4], justify weighted cost models for CU/EU.

Validated energy-complexity frameworks: the ICE model's ability to predict energy from work/span/I/O terms across platforms demonstrates that parameterized energy abstractions can be both analytic and predictive. Our EU dimension extends this approach to explicit $CO_2$ and $ metrics.

Gas-like execution economics: in EVM, opcodes have fixed/dynamic gas costs (e.g., warm vs. cold storage). We generalize this principle to native/IR instruction classes, mapping them to metrics (3) and aggregating them into composite scores for predictable budgeting and trade-off analysis.

## 5. Discussion

### Positioning and novelty

Our work introduces a calibrated, instruction-class, multi-metric model for compute units (CU), energy (EU), carbon footprint ($CO_2$), and monetary cost ($), operating directly at LLVM IR and PTX levels. This approach combines: profile-aware composite scoring to align with performance, cost, or ESG priorities; cross-architecture validation demonstrating predictive agreement with measured runtime and energy; explicit $CO_2$ and $ channels, extending beyond prior energy-only frameworks.

Compared to ICE-style energy complexity, we provide static IR/PTX-level accounting and explicit monetization. Compared to EVM gas schedules, we generalize from bytecode economics to native code and multi-objective trade-offs. Compared to Big O/RAM models, we capture architecture-driven heterogeneity verified through measured instruction latencies and throughputs. The novelty lies in bridging analyzable static models and real-world calibrated data is producing reproducible, profile-aware composite scores that are both theoretically grounded and operationally actionable.

### Feasibility and practical significance

Feasibility is underpinned by three well-established facts: instruction-level cost disparities in cycles and energy are large and consistent across literature; platform-parameterized energy models like ICE are analytically tractable and empirically validated on mainstream CPUs; gas-like costing systems have operationalized per-instruction economics at global scale.

Practical significance includes: architecture-aware algorithm selection in data center, mobile, and HPC contexts; static budgeting of EU/$CO_2$/$ pre-deployment using regional carbon and energy prices; CI regression gates on composite scores to prevent energy or cost regressions in codebases.

### Limitations and threats to validity

We have the following: portability is the cost tables require per-architecture calibration; naive reuse risks misranking across platforms; memory hierarchy & concurrency are the static counts underrepresent cache/NUMA effects and synchronization; hybrid static-dynamic calibration is advisable; dynamic behavior is the branch mispredictions, vectorization, and compiler optimizations can shift realized costs; periodic hardware-grounded recalibration is recommended. ICE comparisons depend on mapping IR counts to work/span/I/O, which may diverge in detail; gas-like mapping is used here as an economic baseline, not a normative schedule.

This comprehensive validation roadmap (Fig. 11) chart illustrates the three-phase approach to maintaining credibility and reproducibility in performance prediction systems:

Phase 1 (Calibration): per-architecture parameter fitting using industry-standard profiling tools, establishing confidence intervals for reliable predictions.

Phase 2 (Validation): direct comparison of predictions against measured performance metrics, ensuring accuracy across time and energy domains.

Phase 3 (Correlation analysis): statistical validation using Spearman and Kendall correlations to verify ranking agreement, providing robust evidence of prediction quality.

The feedback loop ensures continuous improvement, while the monitoring framework maintains long-term system reliability across evolving hardware architectures.

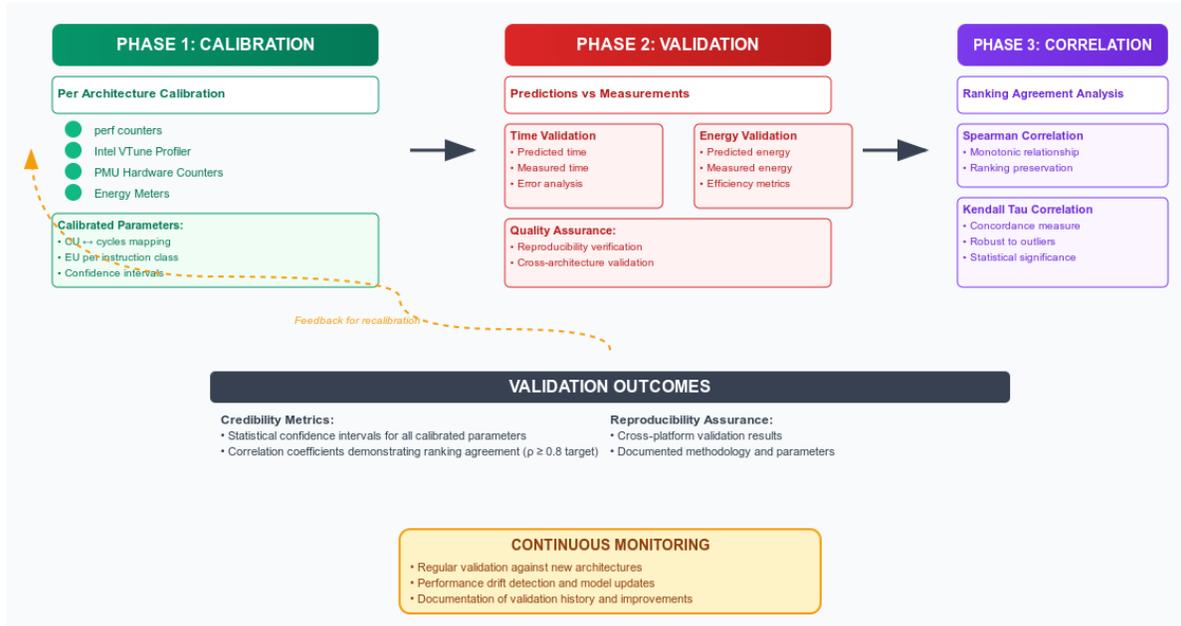

Fig. 11. Validation roadmap for performance prediction

This comprehensive framework chart (Fig. 12) illustrates the 5-pillar approach to cost table versioning and reproducibility:

Central Hub: Semantic versioning system (vMAJOR.MINOR.PATCH) provides structured evolution tracking for architecture-specific cost tables.

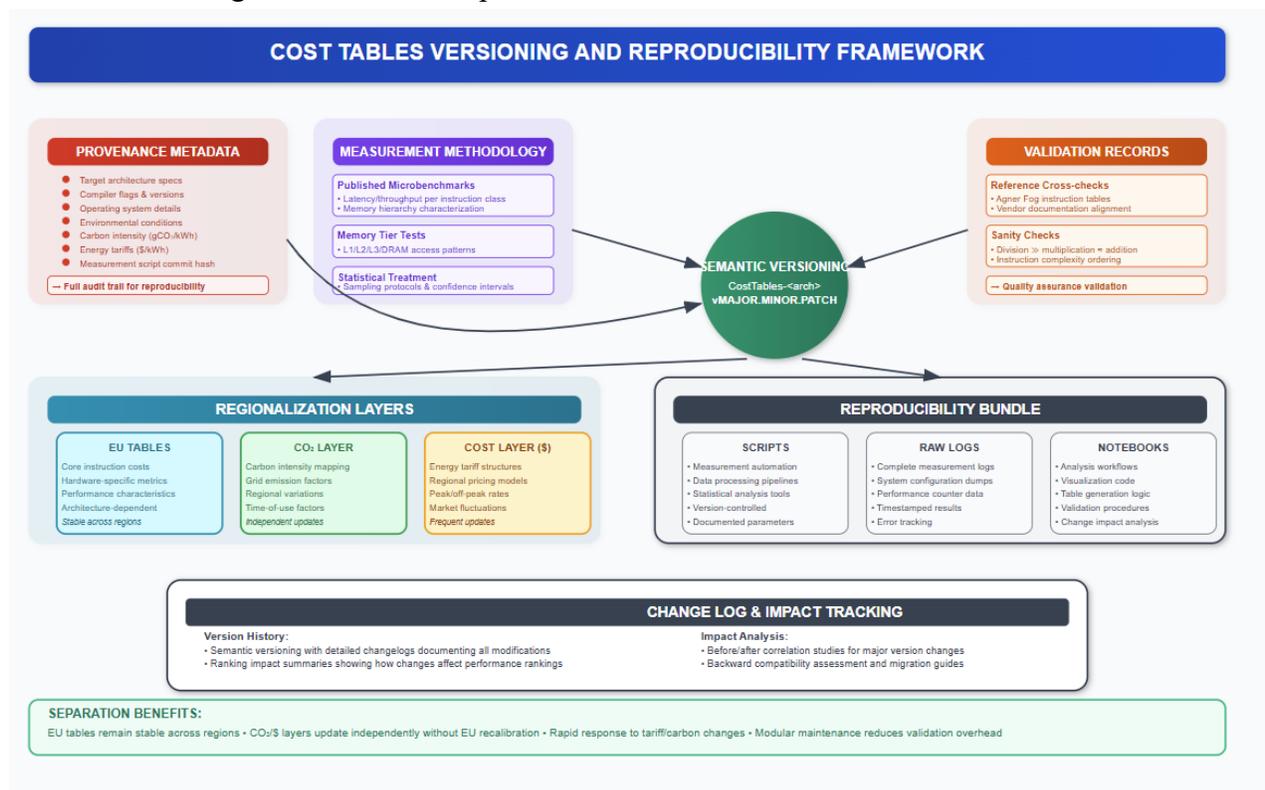

Fig. 12. Comprehensive framework chart with semantic versioning and full traceability

Provenance Metadata: Complete audit trail capturing all environmental factors, from hardware specifications to energy economics, ensuring full traceability.

Measurement Methodology: Standardized protocols using published microbenchmarks and statistical treatments guarantee consistent, comparable results across versions.

Validation Records: Cross-validation against authoritative sources (Agner Fog tables) and logical sanity checks maintain accuracy and credibility.

Regionalization Layers: Architectural separation allows independent updates of carbon intensity and energy tariffs without requiring expensive EU recalibration, enabling rapid response to changing environmental and economic conditions.

Reproducibility Bundle: Complete package of scripts, raw data, and analysis notebooks ensures any version can be fully regenerated, supporting scientific reproducibility and transparency.

The modular design significantly reduces maintenance overhead while maintaining rigorous quality standards across all cost table versions.

By converting EU to $CO_2$ and then monetizing, our framework supports ESG-aligned decisions such as emissions budgeting and SBTi-compliant reductions. Analogous to EVM's pricing deterrents for costly storage, high $CO_2/\$$ weights penalize energy-intensive operations (e.g., division-heavy or memory-bound code under cold/miss-heavy patterns). At the same time, organizations can: profile in situ with calibrated tables and enforce $CO_2$ limits in CI/CD pipelines; time/location shift execution to regions or periods with lower carbon intensity; redesign algorithms to reduce division/mod usage and optimize memory traffic via cache-friendly patterns. Governance benefits include maintaining auditable records linking code changes to $CO_2/\$$ impacts and providing transparent provenance for ESG audits.

## 6. Conclusion

We have introduced a weighted-operation, multi-metric complexity model that unifies four harmonized dimensions: computational units (CU), energy (EU), carbon footprint ($CO_2$), and monetary cost ($\$$) into a profile-driven composite score. By explicitly capturing instruction-level heterogeneity and integrating architecture-specific calibration, our model bridges the gap between theoretical asymptotics and practical, context-aware decision-making.

The approach complements Big-O by preserving its scalability insights while enabling nuanced architecture- and platform-aware comparisons. It aligns naturally with validated energy-complexity theory and extends gas-like execution economics from blockchain VMs to general-purpose native and IR-level code. By embedding $CO_2$ and $\$$ channels alongside CU and EU, the model supports ESG-driven and cost-sensitive decision workflows — something absent from existing complexity frameworks. Our open-source toolchain performs static IR/PTX-level analysis with per-instruction cost mapping, profile-aware scoring, normalization, and grading. At the repository scale, this yields actionable differentiation between algorithms, kernels, and entire code files, uncovering efficiency gaps that asymptotics alone cannot reveal. The methodology is reproducible, parameterizable, and suitable for integration into CI/CD pipelines [21], pre-deployment cost estimation, and energy/carbon budget enforcement.

Key contributions include:
- A general-purpose, calibrated cost model linking instruction classes to CU/EU/$CO_2$/$ across architectures.
- Profile-driven aggregation enabling multi-objective trade-off analysis (performance vs energy vs cost vs sustainability).
- A transparent, versioned calibration protocol for reproducibility and long-term comparability.
- Demonstrated repository-wide analysis showing substantial rank reordering relative to traditional baselines.

Future work will focus on:
- Cross-architecture formal validation on a larger diversity of CPUs, GPUs, and accelerators, including ARM and RISC-V.
- Building standardized calibration datasets and microbenchmark suites for public reproducibility.
- Extending support for dynamic/runtime effects (e.g., branch prediction, vectorization) via hybrid static–dynamic cost modeling.
- Compiler integration to enable cost-aware optimizations at build time.
- Developing predictive models leveraging ML to refine CU/EU/$CO_2$/$ estimates from high-level code patterns without full IR extraction.

By moving from purely theoretical asymptotics toward a reproducible, multi-metric composite metric, this work establishes a practical foundation for cost-aware, architecture-conscious software design – equally relevant to HPC optimization, mobile energy efficiency, and sustainable computing.

**References.**